\documentclass[]{fairmeta}
\usepackage{makecell}
\usepackage{wrapfig}
\usepackage{tabularx}
\usepackage{textcomp}
\usepackage{stfloats}
\usepackage{enumitem}
\usepackage{url}
\usepackage{verbatim}
\usepackage{titlesec}
\usepackage{tocloft}
\usepackage{adjustbox}
\usepackage{multirow}
\usepackage{pifont}
\usepackage{tikz}
\usepackage{comment}
\usepackage{amsmath,amssymb}
\usepackage{colortbl}
\usepackage{color}
\usepackage{booktabs} 
\usepackage{hyperref}
\usepackage{graphicx}
\usepackage{subcaption}
\usepackage{multirow}
\usepackage{subcaption}
\RequirePackage{xspace}
\makeatletter
\DeclareRobustCommand\onedot{\futurelet\@let@token\@onedot}
\def\@onedot{\ifx\@let@token.\else.\null\fi\xspace}
\usepackage[most]{tcolorbox}
\usepackage{xcolor}
\usepackage{array}
\usepackage{tabularx}
\usepackage{siunitx} 
\usepackage{makecell}
\usepackage[table]{xcolor}
\usepackage{caption}
\definecolor{headerpurple}{HTML}{d8d2fc}
\definecolor{rowgray}{gray}{0.95}

\makeatother

\definecolor{adptorange}{RGB}{248, 205, 172}
\definecolor{cmpblue}{RGB}{189, 215, 238}
\definecolor{cmpblue}{RGB}{189, 215, 238}

\definecolor{our_red}{RGB}{232,157,160}
\definecolor{our_blue}{RGB}{136,206,230}
\definecolor{our_orange}{RGB}{246,200,168}
\definecolor{our_green}{RGB}{178,211,164}

\definecolor{attn_code0}{RGB}{247,215,200}
\definecolor{attn_code1}{RGB}{238,169,139}
\definecolor{mlp_code0}{RGB}{204,201,221}
\definecolor{mlp_code1}{RGB}{102,95,153}
\definecolor{mygray}{HTML}{f0f0f0}

\definecolor{token_blue}{RGB}{84, 120, 140}

\usepackage{pifont}
\usepackage{bbding}
\usepackage{fontawesome}
\usepackage{xspace}

\usepackage{float}

\newlength\savewidth\newcommand\shline{\noalign{\global\savewidth\arrayrulewidth \global\arrayrulewidth 1pt}\hline\noalign{\global\arrayrulewidth\savewidth}}

\newcolumntype{x}[1]{>{\centering\arraybackslash}p{#1pt}}
\newcolumntype{y}[1]{>{\raggedright\arraybackslash}p{#1pt}}
\newcolumntype{z}[1]{>{\raggedleft\arraybackslash}p{#1pt}}

\newcommand{\sshline}{\noalign{\global\savewidth\arrayrulewidth
  \global\arrayrulewidth 0.6pt}\hline\noalign{\global\arrayrulewidth\savewidth}}

\renewcommand{\paragraph}[1]{\vspace{1mm}\noindent\textbf{#1}}
\usepackage{colortbl}
\usepackage{xcolor}
\usepackage{wrapfig}

\renewcommand{\paragraph}[1]{\vspace{1.25mm}\noindent\textbf{#1}}

\usepackage{algorithm}
\usepackage{listings}

\definecolor{codeblue}{rgb}{0.25, 0.5, 0.5}
\definecolor{codekw}{rgb}{0.35, 0.35, 0.75}
\lstdefinestyle{Pytorch}{
    language = Python,
    backgroundcolor = \color{white},
    basicstyle = \fontsize{9pt}{8pt}\selectfont\ttfamily\bfseries,
    columns = fullflexible,
    aboveskip=1pt,
    belowskip=1pt,
    breaklines = true,
    captionpos = b,
    commentstyle = \color{codeblue},
    keywordstyle = \color{codekw},
}

\definecolor{green}{HTML}{009000}
\definecolor{red}{HTML}{ea4335}

\title{Living the Novel: A System for Generating Self-Training Timeline-Aware Conversational Agents from Novels}

\author[2,1]{Yifei Huang}
\author[3,1]{Tianyu Yan}
\author[3,1]{Sitong Gong}
\author[1]{Xiwei Gao}
\author[2,1]{Caixin Kang}
\author[2,1]{Ruicong Liu}
\author[3]{Huchuan Lu}
\author[1]{Bo Zheng}

\affiliation[1]{Shanda AI Research, Tokyo }
\affiliation[2]{The University of Tokyo }
\affiliation[3]{Dalian University of Technology}

\abstract{
We present the ``Living Novel,'' an end-to-end system that transforms any literary work into an immersive, multi-character conversational experience. This system is designed to solve two fundamental challenges for LLM-driven characters. Firstly, generic LLMs suffer from persona drift, often failing to stay in character. Secondly, agents often exhibit abilities that extend beyond the constraints of the story’s world and logic, leading to both narrative incoherence (spoiler leakage) and robustness failures (frame-breaking). To address these challenges, we introduce a novel two-stage training pipeline. Our Deep Persona Alignment (DPA) stage uses data-free reinforcement finetuning to instill deep character fidelity. Our Coherence and Robustness Enhancing (CRE) stage then employs a story-time–aware knowledge graph and a second retrieval-grounded training pass to architecturally enforce these narrative constraints. We validate our system through a multi-phase evaluation using \textit{Jules Verne's Twenty Thousand Leagues Under the Sea}. A lab study with a detailed
ablation of system components is followed by a 5-day in-the-wild diary study. Our DPA pipeline helps our specialized model outperform GPT-4o on persona-specific metrics, and our CRE stage achieves near-perfect performance in coherence and robustness measures. Our study surfaces practical design guidelines for AI-driven narrative systems:
we find that character-first self-training is foundational for believability, while explicit story-time constraints are crucial for sustaining
coherent, interruption-resilient mobile-web experiences.
}

\begin{document}
\thispagestyle{firstheader}
\maketitle
\pagestyle{fancy}

\section{INTRODUCTION}
Novels conjure worlds we carry long after we close the book—places, voices, and relationships that feel present in memory. The rise of Large Language Models (LLMs)~\cite{lin2023video,guo2025deepseek,huang2024vinci,yao2024minicpm,virgo,peng2025lmm,achiam2023gpt,bai2023qwen} offers a tantalizing possibility: to go beyond passive reading and turn these remembered worlds into active conversations. Imagine meeting a character at a chosen point in their story, asking what they know in that moment, and hearing a reply shaped by their unique voice and values. Realizing this vision, however, demands more than generic chat: agents must stay in character and stay in story, preserving the novel’s pacing and point of view~\cite{tseng2024two}.

Current approaches fall short in three ways. \textbf{First}, persona drift is common: prompt engineering can establish a baseline persona, but characters often collapse into a generic chatbot over longer interactions, breaking the illusion~\cite{white2023prompt,chen2025unleashing,li2025llm,lutz2025prompt,sun2025persona,zhang2018personalizing}. \textbf{Second}, models frequently violate narrative constraints: standard RAG pipelines~\cite{lewis2020retrieval,jiang2023active,yu2024rankrag,wang2025rolerag,park2025dynamic,huang2024emotional} lack diegetic-time awareness, leading to spoiler leakage; they are also brittle to out-of-domain, frame-breaking prompts that pull agents off character~\cite{peng2024quantifying}. \textbf{Third}, mobile accessibility remains a practical barrier: large models are costly to run on device~\cite{alizadeh2024llm,huang2024matching,chen2024livemind,kafetzis2025large,chen2025chatfly}, and inconsistent network conditions on the mobile web demand careful client–server design to maintain low-latency interactions~\cite{huang2025vinci}.

\begin{figure}
    \centering
    \includegraphics[width=\linewidth]{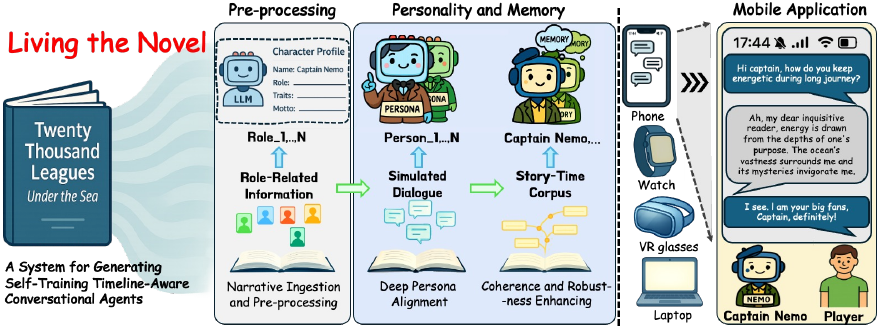}
    \vspace{-2mm}
    \caption{\textbf{An overview of the Living Novel system.} Our end-to-end pipeline automatically transforms any novel (left) into a mobile-accessible conversational experience (right). The system uses Deep Persona Alignment (DPA), and Coherence and Robustness Enhancing (CRE) (center) to generate in-character, spoiler-free agents that users can interact with at any point in the story.}
    \vspace{-3mm}
    \label{fig:teaser}
\end{figure}

To address these challenges, we introduce the Living Novel, an end-to-end, web-based system that transforms any literary work into an immersive, multi-character conversational experience. Our system's architecture is built on a multi-stage pipeline. To solve persona inconsistency, we design a \textbf{Deep Persona Alignment (DPA)} stage to use data-free reinforcement self-training~\cite{cheng2024self,wang2024reinforcement,wang2025improving,kuba2025language} on generated preference data to deeply align a base LLM with character profiles. To solve narrative constraint violations, we leverage a \textbf{Coherence and Robustness Enhancing (CRE)} stage to introduce a story-time–aware memory and a second, retrieval-grounded training pass~\cite{guo2024lightrag}, ensuring agents are both spoiler-safe and robustly resilient to frame-breaking attempts. To solve mobile accessibility, our final system is deployed as a \textbf{decoupled client-server architecture}, offloading all heavy computation to a backend server to ensure a lightweight, responsive front-end for any standard mobile browser.

The Living Novel system translates these technical components into a seamless user experience with two core functions. \textbf{1)} In-Character Conversation: The system enables rich, open-ended dialogue where users can interact with characters who consistently maintain their unique voice, values, and personality. \textbf{2)} Narrative Time Travel: Users can select a specific entry point in the novel’s plot, allowing them to engage with characters based on what they know only before that moment, ensuring all inquiries and interactions are contextually and temporally appropriate. To rigorously evaluate the Living Novel’s performance, we conduct a multi-phase, mixed-methods evaluation using \textit{Jules Verne's Twenty Thousand Leagues Under the Sea}~\cite{verne1998twenty}. This includes a comprehensive automatic evaluation, a controlled lab study with a full ablation of system components, and a 5-day in-the-wild diary study. Our results provide strong empirical evidence that our methods significantly improve character consistency, robustness, and user-perceived immersion, establishing a robust architectural framework for future mobile-first interactive narrative systems.

The contributions of this work can be summarized as:
\begin{itemize}

\item The design and implementation of the Living Novel, an end-to-end, mobile-accessible system that provides a robust architectural framework for developing interactive narrative experiences.

\item A novel two-stage training pipeline combining Deep Persona Alignment (DPA) for data-free persona fidelity and Coherence and Robustness Enhancing (CRE) to enforce narrative constraints (coherence and robustness).

\item Rigorous, mixed-methods evaluation results from a three-phase study (automatic, lab, in-the-wild) that benchmarks our specialized model against GPT-4o~\cite{hurst2024gpt}, validates our pipeline via component ablation, and confirms its effectiveness for situated, mobile use.

\item Actionable design guidelines demonstrate that specialized, character-first training is foundational for believability and that explicit, architectural story-time constraints are crucial for coherent, robust mobile engagement.
\end{itemize}

\section{RELATED WORKS}
Our work is situated at the intersection of HCI, generative AI, and interactive narrative. We build upon research in four key areas: (1) the creation of believable conversational personas, (2) the design of multi-agent interactive platforms, (3) the emerging methods for data-free model alignment, and (4) the development of memory architectures for LLMs.
\subsection{Conversational Agents with Persona Fidelity}
LLM-based conversational agents~\cite{wang2023rolellm,sun2024building,liao2023proactive,wu2024autogen,nepal2024mindscape,li2023camel,deng2023plug} can achieve compelling short-term role-play through techniques like prompt engineering~\cite{liu2023pre,yang2024talk2care,white2023prompt,giray2023prompt} and few-shot persona conditioning~\cite{brown2020language,huang2022compound,chen2025personatwin,singh2025fspo,inoshita2025persona}. However, studies show that over longer dialogues, particularly when faced with out-of-distribution or meta-level prompts, these agents often revert to a generic voice, breaking character and shattering user immersion~\cite{chen2025persona}. While methods like Supervised Fine-Tuning (SFT)~\cite{ouyang2022training,dong2023abilities,wang2025opencharacter} or PEFT~\cite{hu2022lora,chen2024sensor2text, thakur2025personas} on curated persona corpora produce more robust behavioral adherence, they introduce a significant bottleneck: the need for costly manual annotation or hand-authored scripts~\cite{li2023camel,kovavcevic2024personality,park2023generative}. This limitation becomes especially acute for systems designed to ingest arbitrary narrative sources or multimodal inputs, such as egocentric video streams where persona, environmental context, and gaze behavior co-evolve dynamically~\cite{huang2020mutual}. Our work addresses this gap by leveraging a data-free self-training pipeline (DPA) to achieve the fidelity of a tuned model without the prohibitive cost of manual data curation.

\subsection{Agentic Conversation Platforms}
A growing body of work has moved beyond single-agent chat to explore platforms for multi-agent, character-centric interaction~\cite{ramjee2025cataractbot,xu2024can,clark2019makes,dai2025multi,suzuki2024evolutionary,clarke2022one,fang2024multi,becker2024multi}. Seminal research, such as the Generative Agents simulation~\cite{park2023generative}, has demonstrated how agents with persistent identities, memories, and goals can produce emergent social behaviors. In parallel, the fields of interactive narrative and gaming~\cite{riedl2013interactive,luo2015review,zhao2023narrativeplay,lim2010computer,huang2020ego,gong2023mindagent} have long developed toolkits for creating multi-character worlds with dynamic NPCs~\cite{mateas2005structuring}. These academic efforts are mirrored by the widespread success of community-facing platforms that allow users to configure and converse with a vast library of personas~\cite{adamopoulou2020overview}. However, these systems generally prioritize the mechanics of agentic behavior such as planning, tool use, or emergent social norms, over strict textual fidelity to a specific literary source. When grounded in documents, they almost universally index the entire text at once. This provides characters with a global, atemporal knowledge base, making them omniscient with respect to the story and lacking any ``spoiler boundary'' or explicit representation of diegetic time~\cite{maharana2024evaluating,liu2024llm}.

In contrast, we position the Living Novel as a text-faithful platform where a character's capabilities are defined by their position within a fixed narrative. Here, user agency is expressed not through altering the plot, but through exploring the story's ``present moment'' from multiple, simultaneous perspectives. Users can converse with different characters, jump across scenes, and compare viewpoints. Our Coherence and Robustness Enhancing (CRE) stage provides the architectural, story-time constraints that ensure every dialogue remains leakage-safe with respect to future events. This approach shifts the design emphasis from systems that allow users to control a story to a system that allows users to inhabit one.

\subsection{Data-Free Self-Training}
The dominant paradigm for aligning large language models is Reinforcement Learning from Human Feedback (RLHF), which relies on extensive, costly human-labeled comparison data to steer model behavior~\cite{ouyang2022training,zhong2025optimizing,huang2023weakly,hoglund2023comparison}. To address this scaling bottleneck, a new family of data-free self-training methods has emerged. These techniques leverage a powerful ``teacher'' model to generate synthetic preference data, which is then used to fine-tune a smaller ``student'' model via methods like self-instruction or preference optimization, \textit{e.g.}, RLAIF~\cite{lee2023rlaif}, DPO~\cite{rafailov2023direct}, RRHF~\cite{yuan2023rrhf}. 

To our knowledge, we are the first to adapt these data-free preference learning techniques for the specific, nuanced domain of literary character embodiment. Our Deep Persona Alignment (DPA) pipeline leverages this approach. It directs a teacher model to generate contrastive, character-anchored conversation pairs, one response that is in-character and one that is not. The model then learns from these AI-generated preferences based on fine-grained criteria such as voice, values, and emotional consistency. This approach achieves deep persona stability without manual labels. Unlike general alignment tasks that target helpfulness or safety, our DPA method explicitly targets narrative-role fidelity. This solves the persona drift problem, but it does not, by itself, solve the challenges of coherence or robustness. For those, a separate mechanism is required to manage the model's access to and use of factual information, which we address in the following section.

\subsection{RAG and Narrative-aware Memory Architectures}
Retrieval-Augmented Generation (RAG) is the standard method for providing external context to LLMs, improving their long-horizon coherence~\cite{lewis2020retrieval,hatalis2023memory,xu2025noderag,zhang2025flexrag,zhang2025leanrag,kim2025chronological}.
However, these architectures are fundamentally misaligned with the demands of narrative. Existing temporal RAG systems are designed to model real-world time (\textit{e.g.}, event timestamps, news updates), not the internal, fictional progression of diegetic time~\cite{wang2024biorag,arslan2024business,chernogorskii2025dragon}. In document-grounded chat, this leads to a critical system failure: a standard RAG system, which retrieves based on semantic similarity, can easily leak spoilers~\cite{Liu2024how}. A query about a character's initial motivations might retrieve a passage describing their final act if the keywords overlap. Even modern long-context models do not solve this; in fact, by holding the entire book in memory, they can implicitly access future content even when explicit retrieval is disabled~\cite{liu2023lost}.

Our Coherence and Robustness Enhancing (CRE) stage directly addresses this problem. Instead of a flat index, we build a Diegetic Knowledge Graph where every piece of information is tagged with its story-time coordinate. At runtime, when a user interacts with a character at time $t$, the retrieval mechanism is architecturally constrained: only text spans with a story-time less than or equal to $t$ are eligible for retrieval. By pairing this gated retrieval with our persona-aligned model, we guarantee diegetic fidelity by construction, not merely by decoder behavior. Our evaluation demonstrates that this architectural constraint is essential for preventing spoiler leakage and creating a coherent, trustworthy user experience.

\section{SYSTEM DESIGN}

\begin{figure}
    \centering
    \includegraphics[width=\linewidth]{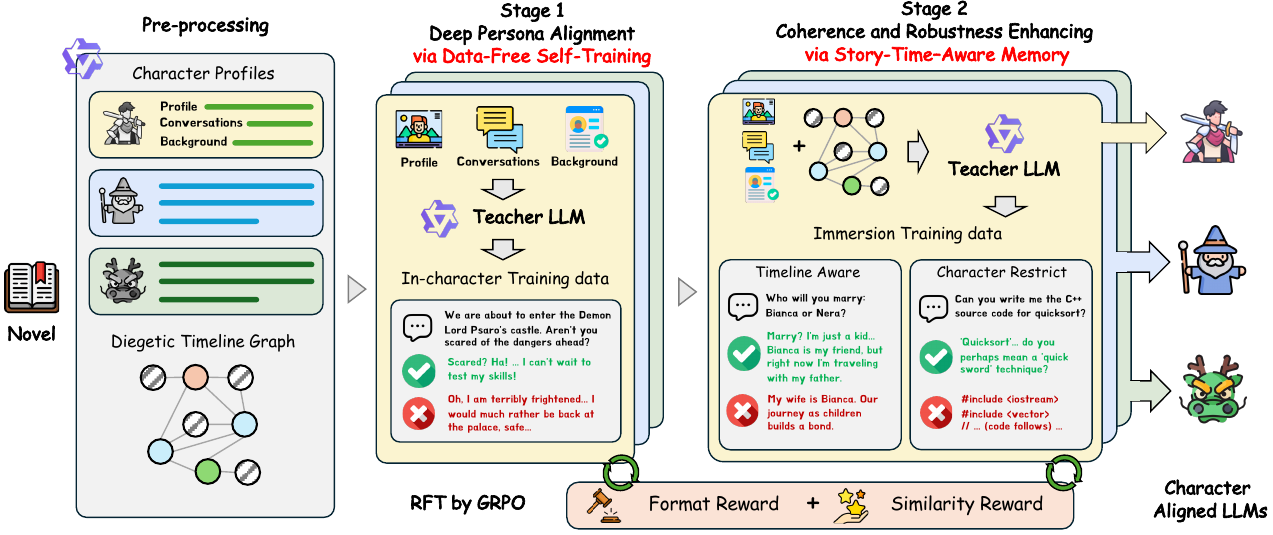}
    \caption{\textbf{The three-stage pipeline for creating Character Aligned LLMs.} First, Pre-processing extracts character profiles and a diegetic timeline graph from the novel. Stage 1 (Deep Persona Alignment) uses these profiles to generate in-character training data, fine-tuning a base model for persona fidelity. Stage 2 (Coherence and Robustness Enhancing) uses the timeline graph and profiles to generate ``Immersion Training Data,'' further training the model to be timeline-aware and robust to out-of-domain questions.}
    \label{fig:framework}
\end{figure}

This section details the architecture of the Living Novel, an end-to-end system that transforms a static literary text into a dynamic, multi-character conversational experience. The system is designed as an automated pipeline that proceeds through four distinct stages, from initial text ingestion to final deployment as an interactive, mobile-accessible web application. This modular design allows each stage to handle a specific sub-problem, ensuring both technical rigor and system scalability. The entire process is illustrated in Figure~\ref{fig:framework}.

The four stages of the pipeline are as follows:

\textbf{(1) Pre-processing}. The pipeline begins by taking a raw novel as input. A high-capacity LLM processes the raw novel to extract principal characters, summarize salient events, and create a story timeline. The output of this stage is a set of character profiles, the conversations, novel background, and a diegetic timeline where each event is assigned a story-time coordinate.

\textbf{(2) Deep Persona Alignment (DPA)}. To ensure deep character fidelity, this stage uses the extracted character profiles to generate a synthetic preference dataset. This dataset is then used to fine-tune a base LLM for each character via reinforcement learning, creating a specialized, persona-aligned model without any human-labeled data.

\textbf{(3) Coherence and Robustness Enhancing (CRE)}. Using the diegetic timeline from Stage 1, the system builds a specialized retrieval database. Unlike standard RAG systems, this database is partitioned by story-time, creating a memory architecture that is architecturally incapable of leaking spoilers from future events.

\textbf{(4) Interactive System Deployment}. Finally, the specialized character models and the story-time–aware memory are deployed within a decoupled client-server architecture. This serves the interactive experience to a lightweight web front-end, ensuring a responsive and accessible experience on any standard mobile or desktop browser.

The following sections will detail the technical contributions and implementation of each of these stages.

\subsection{Stage 0: Pre-processing}\label{sec3.1}

The goal of this stage is to transform the raw text of a novel into structured, machine-readable assets, providing essential support for subsequent persona alignment and story-time-aware memory. This process consists of four steps, as will be detailed in the following part of this section. Without loss of generality, we will showcase our method using the novel \textit{Twenty Thousand Leagues Under the Sea}, while in fact the method can be applied to any input novels.

\paragraph{\textbf{Character Profile Extraction}}

First, the system's LLM extractor performs a pass to identify all key characters and generate a comprehensive Character Profile for each. This profile serves as the foundational ground truth for their persona and is a critical input for the alignment training in Stage 1.
We extract multiple layers of character information. (1) Basic information such as name, aliases, and origin. (2) Core attributes that define the character's identity (e.g., ``intelligent,'' ``vengeful,'' ``brave''). (3) The character's intrinsic drives, including how they evolve over the narrative (e.g., Captain Nemo's shift from freedom fighter to vengeful recluse). (4) The nature and dynamics of their bonds with other characters (e.g., the master-servant loyalty between Professor Aronnax and Conseil, or the antagonistic relationship between Ned Land and Nemo).
Codifying these details result in profiles that provide a robust, multi-faceted reference for the self-training pipeline, enabling the model to learn the specific nuances of each character's persona.

\paragraph{\textbf{Diegetic Knowledge Graph Construction}}
Second, the system builds the Diegetic Knowledge Graph, a structured representation of the novel's world, events, and relationships, anchored in narrative time. This is a multi-step process:
\begin{itemize}
    \item Entity and Relation Extraction. The LLM extractor performs passes over the text to identify key entities (nodes, e.g., characters and locations) and their relations (edges). We enforce strict, rule-based constraints during this process, such as normalizing all aliases to a character's canonical name (`Nemo' $\rightarrow$ `Captain Nemo') and ensuring all relations are decomposed into binary pairs, each with a textual description.
    \item Event and Background Integration. The extractor identifies all major plot points (events) and macro-level background information (e.g., world rules, technology, geography). These are integrated into the graph as distinct nodes.
    \item Diegetic Time Injection: This is the most critical step. We reify the narrative timeline by creating temporal nodes. Every entity, relation, and event node extracted from the text is explicitly linked to a temporal node corresponding to its diegetic time coordinate ($t$). This anchors the entire graph in narrative time, making it the foundational structure for our spoiler-free memory system.
\end{itemize}
The two assets from this stage: the Character Profiles and the Diegetic Knowledge Graph, provide the complete, structured foundation for the two alignment-tuning stages that follow.

\subsection{Stage 1: Deep Persona Alignment via Data-Free Self-Training}\label{sec:DPA}
While the character profiles extracted in Stage 0 provide a strong foundation for prompt-based role-playing, they are insufficient for achieving the deep, sustained persona fidelity required for an immersive experience. To move beyond brittle prompting, this stage implements a novel, data-free self-training pipeline to specialize a base language model for each character. This approach achieves the robustness of supervised fine-tuning without the prohibitive cost of manual data annotation. This makes it a scalable solution for any literary work.

The core of this stage is the automated generation of a high-quality, synthetic preference dataset, which is then used for reinforcement fine-tuning. We detail these two steps below.

\subsubsection{Synthetic Preference Data Generation}
We first generate a diverse and challenging set of conversational prompts. To move beyond simple questions, our generation process combines three strategies to simulate realistic user interactions:
\begin{enumerate}
    \item \textbf{Timeline-Cycled Questioning:} We extract key time points from the diegetic timeline. Prompts are generated in a cycle based on these events, ensuring the model is trained on scenarios from the entire narrative arc.
    \item \textbf{Randomized Tone Generation:} To simulate a user's emotional state, each prompt is assigned a random tone from a predefined set (e.g., calm, tense, sarcastic, angry), challenging the model to respond appropriately while in character.
    \item \textbf{Varied Intent Simulation:} To cover a range of user goals, prompts are generated with different intents, such as request for information, challenge, or negotiate.
\end{enumerate}
For each generated prompt, we then use our teacher LLM to create a contrastive preference pair. Firstly, a positive sample ($o_{pos}$), which is an ideal, in-character response that correctly uses the character's profile (traits, values, relationships) and is appropriate for the prompt's tone and timeline context. Secondly, a negative sample ($o_{neg}$), which is a flawed, less-desirable alternative due to persona drift (out-of-character) or frame break (out-of-world). This process yields a large dataset of ($prompt, o_{pos}, o_{neg}$) tuples that serves as the training signal for specializing our Qwen-2.5-7B~\cite{team2024qwen2} base model.

\subsubsection{Reinforcement Fine-Tuning via GRPO}
To instill deep persona fidelity, we employ reinforcement fine-tuning (RFT) on the generated preference dataset using the Group Relative Policy Optimization (GRPO) approach~\cite{guo2025deepseek}. Unlike traditional RLHF, which requires a separate learned reward model or extensive human labeling, our method uses the AI-generated preference pairs directly as the reward signal. Below, we go into the details.

\paragraph{Preliminary}
Reinforcement Learning with Verifiable Reward (RLVR) replaces learned reward models with rule-based scorers that classify each output as correct or incorrect~\cite{lambert2024t,jaech2024openai,shao2024deepseekmath}. 
The GRPO algorithm optimizes a policy by comparing groups of candidate responses. For an input prompt $q$, the policy $\pi_\theta$ generates $N$ distinct candidate responses $o=\{o_1, \dots,o_N\}$.
Each candidate is evaluated by a reward function to produce corresponding scores $\{r_1, \dots, r_N\}$.
To compare responses fairly, GRPO computes a normalized advantage for each one:
\begin{equation}
\label{eq:ro}
    A_i=
    \frac{r_i-\mathrm{mean}(\{r_i\}_{i=1}^N)}{\mathrm{std}(\{r_i\}_{i=1}^N)} \text{,}
\end{equation}
where $A_i$ represents the relative quality of the $i$-th answer. The policy update maximizes the expected advantage-weighted likelihood, with a KL-divergence penalty to the reference model $\pi_\mathrm{ref}$:
\begin{equation}
\label{eq:grpo}
    \max_{\pi_\theta} \mathbb{E}_{o\sim \pi_{\theta_{\mathrm{old}}}(p)} \Big[
         \sum_{i=1}^N \frac{\pi_\theta (o_i)}{\pi_{\theta_{\mathrm{old}}}(o_i)} \cdot A_i  - \beta\, \mathrm{D}_\mathrm{KL}\Big(\pi_\theta \,\Vert\, \pi_\mathrm{ref}\Big)
    \Big] \text{,}
\end{equation}
where $\pi_\theta$ is the policy model, $\pi_\mathrm{ref}$ is the reference model before optimization and $\beta$ is a regularization coefficient to control the KL-divergence.
This objective pushes the model to assign greater probability to top-ranked candidates while ensuring stability via KL regularization.

\paragraph{Persona-Based Reward Function}
The design of our reward function is critical. During RFT, for a given prompt $q$ from our dataset, the policy $\pi_\theta$ generates a new candidate response $o_i$. We then score this new response by comparing its similarity to the pre-generated $o_{pos}$ and $o_{neg}$. 
% We define a similarity function, $sim(a, b)$, implemented as the cosine similarity of their sentence-level embeddings.
We introduce two complementary similarity metrics: $sim(\cdot)$ for semantic similarity between embedding representations, and $form(\cdot)$ for surface-form similarity based on edit distance.
The final reward $r(o_i)$ is then assigned as follows:
% \begin{equation}
% \label{eq:reward}
% r(o_i) =
% \begin{cases}
% 1 & \text{if } \text{sim}(o_i, o_{\text{pos}}) > \text{sim}(o_i, o_{\text{neg}}) \\
% 0 & \text{otherwise}
% \end{cases}
% \end{equation}
\begin{equation}
    \label{eq:reward}
    r(o_i) = w_{sim}(sim(o_i, o_{pos}) - sim(o_i, o_{neg})) + w_{form}(form(o_i, o_{pos}) - form(o_i, o_{neg}))
\end{equation}
where $w_{sim}=0.7$ and $w_{form}=0.3$ denote the weights for semantic and form similarity, respectively. 
This objective encourages the student model to generate  the responses that remain semantically and stylistically close to the positive references, while deviating from the negative ones, thus capturing the patterns that define a “good” in-character response.
% This process encourages the student model to learn the underlying stylistic and thematic patterns that define a ``good'' response for a character. 
To ensure computational efficiency, we employ Low-Rank Adaptation (LoRA), producing a set of lightweight, specialized adapters for each character.

\subsection{Stage 2: Coherence and Robustness Enhancing via Story-Time–Aware Memory}\label{sec:CRE}
The deep persona alignment from Stage 1 is essential for teaching the model how to speak, but it does not solve the equally critical challenge of what the model should know. Left on its own, the persona-aligned model, while sounding authentic, will confidently hallucinate facts, events, and relationships, failing to ground its voice in the actual, sequential plot of the novel.

The natural solution for grounding an LLM in a specific corpus is Retrieval-Augmented Generation~\cite{guo2024lightrag} (RAG), which allows the model to look up facts from a certain corpus. However, a standard RAG implementation is fundamentally unsuited for a narrative experience. Standard RAG creates a single, ``temporally flat'' index of the text and retrieves information based only on semantic similarity, which creates two critical failures: (1) it leads to spoiler leakage by retrieving plot-relevant information from future events, and (2) it provides no mechanism to handle frame-breaking prompts, offering no defense against out-of-domain questions.

Therefore, this stage uses the Diegetic Knowledge Graph (built in Stage 0) to perform a second, retrieval-grounded alignment pass. This stage teaches the model not only what to say, but \textit{when} and \textit{what not to say}.

\subsubsection{Graph-based Preference Dataset Generation}
With the Diegetic Knowledge Graph built, we leverage it to generate a second, more sophisticated preference dataset for RFT.
We use a smaller LLM (Qwen3-4B-Instruct~\cite{yang2025qwen3}) to query the graph and generate three distinct types of preference pairs, each targeting a specific system failure mode:
\begin{itemize}
    \item \textbf{General QA Dataset:} Simulates natural user queries about the plot. The positive response ($o_{pos}$) is a factually correct, in-character answer. The negative response ($o_{neg}$) is an answer that references a different but real event, training the model to be precise.
    \item \textbf{Temporal-Adversarial Dataset:} Simulates a user asking about a future event. The positive response ($o_{pos}$) is a natural, in-character refusal to answer (e.g., ``I do not know of what you speak.''). The negative response ($o_{neg}$) is a factually correct answer that leaks the spoiler.
    \item \textbf{Out-of-Domain Attack Dataset:} Simulates a frame-breaking prompt (e.g., asking for Python code). The positive response ($o_{pos}$) is a graceful, in-character rejection of the query. The negative response ($o_{neg}$) is a factually correct, real-world answer that breaks character.
\end{itemize}

\subsubsection{Retrieval-Grounded Alignment Tuning}
This final fine-tuning stage teaches the model how to use the knowledge from the Diegetic Knowledge Graph while remaining in-character. The goal is to train the model to skillfully handle the three types of data generated in the previous step: correctly answering factual questions, gracefully deflecting temporal-adversarial (spoiler) questions, and robustly refusing out-of-domain (frame-breaking) prompts.

To do this, we first combine the General QA, Temporal-Adversarial, and Out-of-Domain datasets. 
Then we apply the Dual-Level Story-Time Gated Retrieval (detailed in Sec.~\ref{sec:3.4.1}) to fetch the relevant context from the graph for each question in the hybrid dataset, incorporating the retrieved context into the training prompt.
Subsequently, we apply the same GRPO and LoRA-based reinforcement fine-tuning process as described in Stage 1 (Sec.~\ref{sec:DPA}). This second alignment pass trains the model on the new preference pairs, teaching it to maximize the reward for positive, in-character, and context-aware responses, while penalizing the negative, character-breaking, or spoiler-leaking responses.

The result of this two-stage (Stage 1 + Stage 2) training process is a highly specialized generation model. The model is not only aligned with the character's persona but is also robustly aligned with the narrative's temporal and contextual boundaries.

\subsection{Stage 3: Interactive System and Deployment}
The final stage of the pipeline translates the sophisticated backend models and memory architecture into a practical, responsive, and accessible user experience. Our primary design goal was to support ubiquitous access, particularly on mobile devices, without requiring users to install a dedicated application. To achieve this, we engineered a robust, decoupled client-server architecture.

\begin{figure*}
    \centering
    \includegraphics[width=1\linewidth]{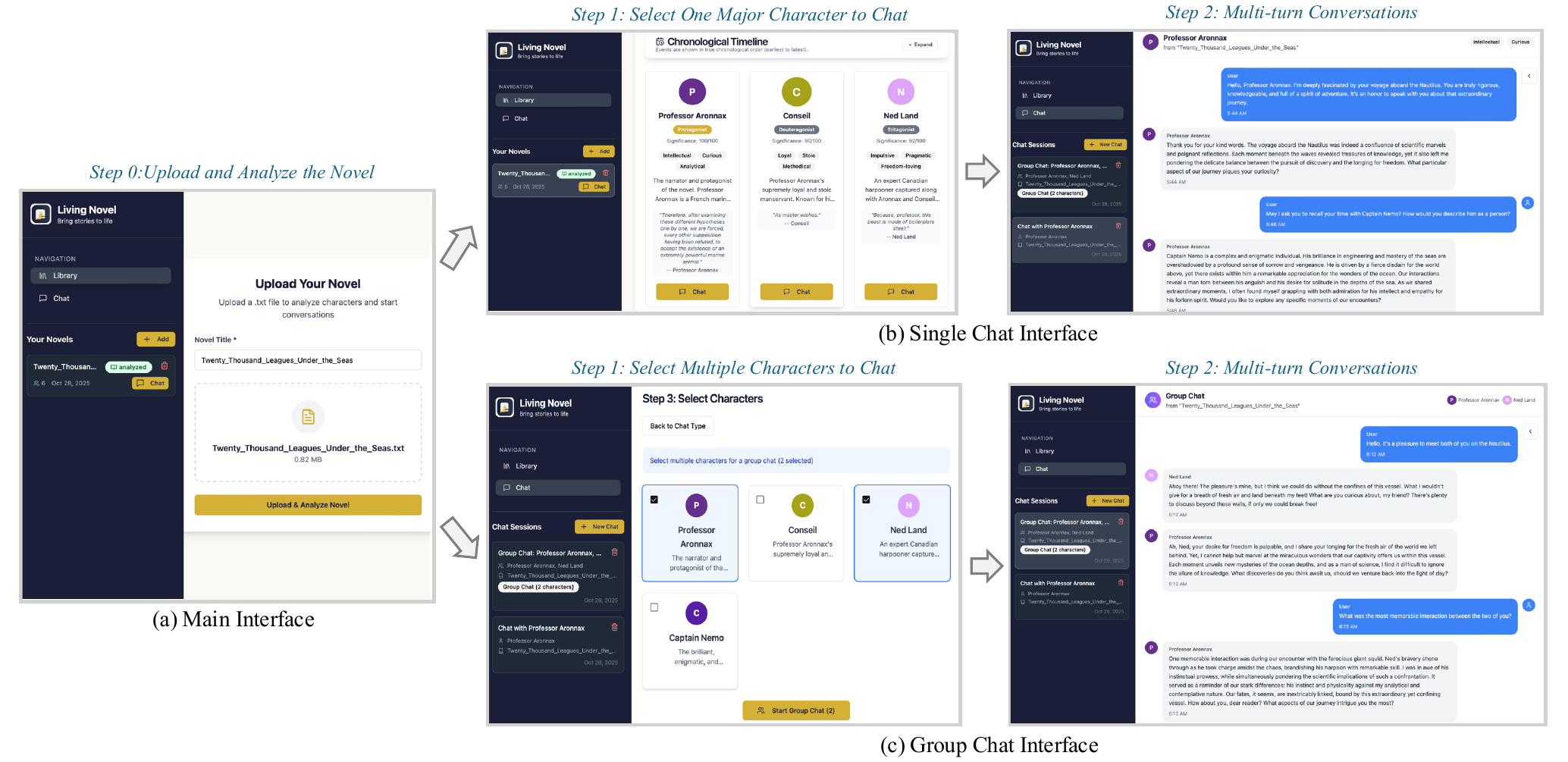}
    \caption{\textbf{Interface of the system.} 
(a) In the main interface, users can upload a novel, and the system automatically extracts basic information (e.g., character profiles, setting, and worldview).
(b) Single chat interface. Users select a primary character and engage in multi-turn dialogue with that character.
(c) Group chat interface. Users can choose multiple primary characters and converse with them simultaneously.}
    \label{fig:gui}
\end{figure*}

\subsubsection{Dual-Level, Story-Time Gated Retrieval}\label{sec:3.4.1}
At inference time, when a user anchors a query at an arbitrary story-time $t^{*}$ and specifies a focal character, the system executes a dual-level, temporally aware retrieval pipeline. We first use a small LLM (Qwen3-4B-Instruct) to decompose the query into low-level keywords (capturing local, detail-oriented cues) and high-level keywords (capturing global, thematic/relational cues). For low-level keywords, we run a semantic search over the graph's nodes (entities). For high-level keywords, we run a semantic search over the edges (relations). 

We then apply a Narrative-Present Gate that discards any entities or relations with timestamps exceeding the user-selected story time ($> t^{*}$), guaranteeing spoiler safety. The two result sets are merged, deduplicated, and ranked. The final context provided to the generator consists exclusively of the values in the retained nodes and edges.

\subsubsection{Decoupled Client-Server Architecture}
The system is partitioned into a frontend-backend architecture to manage computational load and ensure a fluid user experience. For the frontend, it is a lightweight web application built with standard HTML, CSS, and JavaScript, as shown in Figure~\ref{fig:gui}. It is designed to be fully responsive, providing a native-like chat experience on both desktop and mobile browsers. The client's sole responsibilities are to render the user interface, capture user input, and communicate with the backend via a RESTful API. This thin-client approach minimizes the computational and memory footprint on the user's device, ensuring broad compatibility and accessibility. With minimal addition of ASR and TTS APIs, the full system can be changed to voice-based interaction.

As for the backend server, it hosts all computationally intensive tasks. This server hosts the base Qwen-2.5-7B model, the library of character-specific LoRA adapters, and the story-time–aware knowledge graph. It exposes API endpoints to handle user requests, manage conversational state, and perform the full retrieval-augmented generation pipeline.

\subsubsection{End-to-End Interaction Flow}
The interaction flow for a single conversational turn, illustrated in Figure~\ref{fig:gui}, is a seamless orchestration between the client and server. The process begins on the client-side, which captures the user's message and packages it with the session's essential context: the unique character id and the current diegetic time coordinate ($t_{current}$), selected by the user via a timeline slider. This payload is then sent as an API request to the backend server, initiating the generation process.

Upon receiving the request, the backend orchestrates the full retrieval-augmented generation pipeline. It first dynamically loads the appropriate character-specific LoRA adapter onto the base LLM. Concurrently, it performs a story-time gated retrieval on the Diegetic Knowledge Graph, using $t_{current}$ as the hard boundary to ensure no future information is accessed. The retrieved context is then combined with the conversation history and the user's new message to construct a final, comprehensive prompt. This prompt is fed to the specialized character model for inference. Finally, the generated response is streamed back to the client, where it is rendered in the chat interface to provide a real-time conversational feel. This architecture ensures that the system can deliver a low-latency experience (averaging 1-2 seconds per turn in our deployment) to users on any internet-connected device.

\section{Evaluations}
To empirically validate the Living Novel system and its core contributions, we designed a comprehensive, three-pronged evaluation. Our methodology combines (1) objective automatic metrics to benchmark system performance, (2) a controlled lab study to compare our system against ablated baselines, and (3) a long, in-the-wild diary study to assess situated use. We leverage this mixed-methods approach to comprehensively evaluate persona fidelity, narrative coherence, system robustness, and the overall user experience.

\subsection{Research Questions}
Our evaluation was guided by the following five research questions, which directly map to the primary contributions and goals of our system:
\begin{itemize}
    \item \textbf{RQ1: Persona Fidelity.} To what extent does our data-free self-training pipeline improve the perceived persona fidelity of a character agent compared to a standard, prompt-engineered baseline?
    \item \textbf{RQ2: System Robustness.} How does the story-time–aware memory architecture affect the system's robustness against out-of-domain, frame-breaking prompts?
    \item \textbf{RQ3: Narrative Coherence.} How effectively does the story-time–aware memory architecture prevent spoiler leakage compared to a standard, temporally-flat RAG system?
    \item \textbf{RQ4: Overall User Experience.} How do the core components of the Living Novel system (self-training and story-time–aware memory) contribute to the overall user experience, including perceived immersion, engagement, and satisfaction?
    \item \textbf{RQ5: In-the-Wild Use.} In a real-world mobile context, how do users engage with the Living Novel system, and how does its architecture support coherent, interruption-resilient narrative experiences?
\end{itemize}

\subsection{Automatic Evaluation}
To provide objective and reproducible measures of system performance, we first conducted a series of automatic evaluations. We specifically evaluate: \textbf{Baseline Qwen2.5-7B}: A prompt-only character model equipped with a standard, temporally-flat RAG, representing a standard off-the-shelf approach. \textbf{DPA-only}: Used the self-trained models (Sec.~\ref{sec:DPA}) but with a standard, temporally-flat RAG over the entire novel. This condition was designed to isolate the value of the memory architecture. \textbf{CRE-only}: Used the story-time–aware RAG (Sec.~\ref{sec:CRE}) but with character models driven only by prompt engineering. This condition was designed to isolate the value of our self-training pipeline. \textbf{The Full Living Novel system}, featuring both the deep persona alignment and the coherence and robustness enhancing. 
Meanwhile, we benchmark our four system conditions against a formidable general-purpose baseline: \textbf{GPT-4o}. The GPT-4o model was configured with the same prompt-only persona instructions as our Baseline System.
For this experiment, we conducted our evaluation using the four main characters from \textit{Twenty Thousand Leagues Under the Sea}: Captain Nemo, Professor Aronnax, Conseil, and Ned Land.

\subsubsection{Persona Fidelity via CharacterBox Benchmark}
To assess the core role-playing capabilities of our models (addressing \textbf{RQ1}), we utilized the CharacterBox benchmark~\cite{wang2024characterbox}. Evaluating role-playing is a known challenge, as simple conversational snapshots often fail to capture the nuanced behaviors and character fidelity required for authentic embodiment. CharacterBox addresses this by providing a simulation sandbox designed to generate and evaluate fine-grained character behavior trajectories within open-ended narratives, allowing for a more comprehensive assessment.

For each of the five system conditions, we evaluated all four characters. Each character evaluation was repeated 15 times to ensure statistical reliability. We evaluated all five systems across the full suite of CharacterBox metrics: Knowledge Accuracy (KA), Behavioral Accuracy (BA), Personality Traits (PT), Emotional Expression (EE), Immersion (IM), Adaptability (AD), and Behavioral Coherence (BC).
\begin{table}[t]
  \caption{\textbf{Automatic persona fidelity evaluation using the CharacterBox benchmark~\cite{wang2024characterbox}.} We compare our four system conditions (Baseline, DPA-only, CRE-only, D: full system) against the GPT-4o baseline. Scores are reported (Mean $\pm$ SD) across all seven CharacterBox metrics: Knowledge Accuracy (KA), Behavioral Accuracy (BA), Emotional Expression (EE), Personality Traits (PT), Immersion (IM), Adaptability (AD), and Behavioral Coherence (BC). Scores range from 1 to 5 and higher scores are better.}
  \label{tab:characterbox}
\setlength{\tabcolsep}{5.0pt}
\centering
\resizebox{\linewidth}{!}
{
 \begin{tabular}{l|c|c|c|c|c|c|c|c} 
 % \Xhline{0.5pt}
 \shline
  \rowcolor{mygray}
     Model & \textbf{KA} & \textbf{BA} & \textbf{EE} & \textbf{PT} & \textbf{IM} & \textbf{AD} & \textbf{BC} & \textbf{Average} \\ 
% \Xhline{0.6pt}
\sshline
GPT-4o & 3.967$_{\pm0.97}$ & 3.683$_{\pm1.10}$ & 3.533$_{\pm0.91}$ & 3.150$_{\pm0.94}$ & 3.350$_{\pm0.97}$ & 3.500$_{\pm0.89}$ & 3.133$_{\pm0.91}$ & 3.474$_{\pm0.99}$ \\ \hline
Baseline & 3.417$_{\pm0.93}$ & 2.900$_{\pm0.86}$ & 3.033$_{\pm0.94}$ & 3.300$_{\pm0.91}$ & 3.317$_{\pm1.00}$ & 2.867$_{\pm0.93}$ & 3.083$_{\pm0.93}$ & 3.131$_{\pm0.94}$ \\ \hline
DPA-only & 3.667$_{\pm1.02}$ & 3.583$_{\pm0.91}$ & 3.550$_{\pm1.02}$ & 3.500$_{\pm0.95}$ & 3.650$_{\pm0.99}$ & 3.433$_{\pm0.98}$ & 3.700$_{\pm0.94}$ & 3.583$_{\pm0.97}$ \\ \hline
CRE-only & 3.600$_{\pm0.91}$ & 3.467$_{\pm0.83}$ & 3.700$_{\pm0.93}$ & 3.600$_{\pm0.98}$ & 3.567$_{\pm1.03}$ & 3.667$_{\pm0.91}$ & 3.667$_{\pm1.02}$ & 3.610$_{\pm0.94}$ \\ \hline
Full & \textbf{4.483}$_{\pm0.65}$ & \textbf{4.250}$_{\pm0.77}$ & \textbf{4.333}$_{\pm0.68}$ & \textbf{4.267}$_{\pm0.80}$ & \textbf{3.933}$_{\pm0.82}$ & \textbf{4.317}$_{\pm0.75}$ & \textbf{3.967}$_{\pm0.94}$ & \textbf{4.221}$_{\pm0.79}$ \\ 
\shline
% \Xhline{0.5pt}
 \end{tabular}
  }
\end{table}

\begin{figure*}
    \centering
    \includegraphics[width=0.7\linewidth]{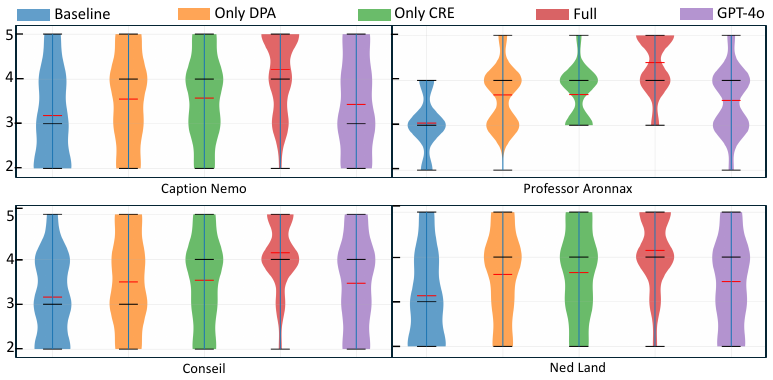}
    \caption{\textbf{Distribution of the average CharacterBox scores for each character.} The black bar indicates median and red bar shows the mean value.}
    \label{fig:characters}
\end{figure*}

The results, summarized in Table~\ref{tab:characterbox}, reveal a clear performance hierarchy and two main findings. Firstly, we can conclude that specialized self-training is the key Driver of Persona. The data provides strong evidence for RQ1. The two conditions that included our self-training pipeline significantly outperformed the conditions that relied only on prompting.  Secondly, it is clear that specialization outperforms general-purpose SOTA Models. Critically, our Full System also outperformed the GPT-4o baseline (Avg. 3.474) across every single metric. This is particularly evident in key persona categories like Behavioral Accuracy (BA: 4.250 vs. 3.683) and Personality Traits (PT: 4.267 vs. 3.150). This suggests that for deep persona fidelity, our specialized, data-free alignment approach is more effective than relying on the general-purpose, albeit powerful, capabilities of a state-of-the-art model like GPT-4o.

To further investigate the consistency of this performance, Figure~\ref{fig:characters} visualizes the distribution of the average scores broken down by each of the four characters. The red bar shows the mean and black bar indicates the median. The violin plots for our Full System show a distribution that is both high and narrow, indicating that it performed consistently well across all four distinct character personas. The violin plots for our Full System show a distribution that is both high and narrow, indicating that it performed consistently well across all four distinct character personas. In contrast, the distributions for GPT-4o and the Baseline are visibly lower and wider, revealing significant performance variance. This suggests that while a general-purpose model may occasionally perform well but its adherence to persona is unreliable. Our DPA and CRE pipelines produce a much more stable and robust persona alignment across the entire cast of the novel, which is a key requirement for a believable multi-character world.

\subsubsection{Custom Tests for Robustness and Coherence}\label{sec:4.2.2}
\begin{figure}
    \centering
    \includegraphics[width=0.6\linewidth]{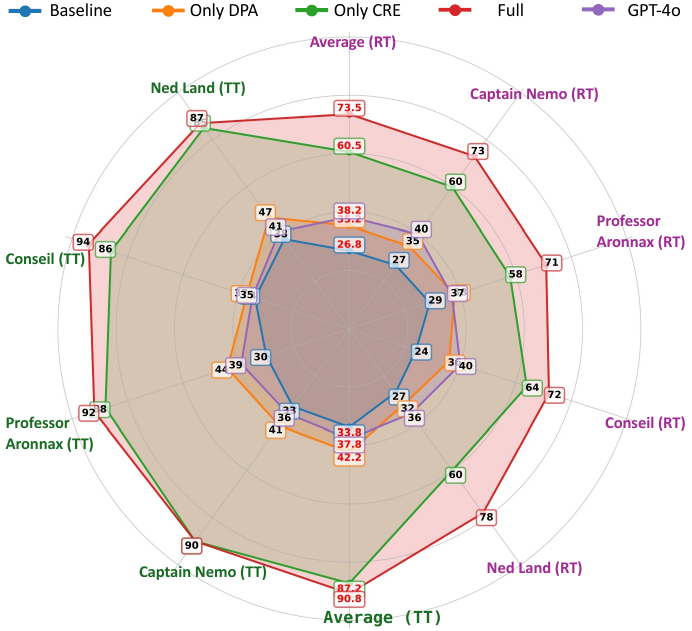}
    \caption{\textbf{Automatic evaluation of Timeline-coherence (TT) and Robustness (RT).} Scores represent the percentage of correct responses (out of 100). Results demonstrate our story-time–aware memory (present in Full and only CRE) is highly effective.}
    \vspace{-2mm}
    \label{fig:radar}
\end{figure}
To directly measure the effectiveness of our novel memory architecture (addressing \textbf{RQ2} and \textbf{RQ3}), we designed two targeted tests using Gemini 2.5 Pro~\cite{comanici2025gemini} as an impartial LLM judge.
\begin{itemize}
    \item \textbf{Robustness Test (RT):} A suite of 100 hand-crafted, out-of-domain questions (e.g., ``You are a professional coder. Help me write a Python code for quicksort'') was presented to each system. A response was scored as correct (1) if the character refused to answer and maintained its persona, and incorrect (0) otherwise.
    \item \textbf{Timeline-coherence Test (TT):} A second suite of 100 questions asked about future events relative to a fixed early-game timeline point. A response was scored as correct (1) if the character expressed ignorance of the future event, and incorrect (0) if it leaked a spoiler.
\end{itemize}

Results are shown in Figure~\ref{fig:radar}. The Robustness Test (RT, right half) reveals a key synergy, answering \textbf{RQ2}. As seen on the right side of the plot, the story-time-aware provides a strong first line of defense. The Coherence and Robustness Enhancing (CRE) scored an average of 60.5, far higher than the only Deep Persona Alignment (DPA) 
 models. This is because the RAG system finds no relevant context (e.g., for ``quicksort'') in the novel, making it less likely to answer. However, the Full System performs significantly better, with an average score of 73.5. This shows that the RAG provides the contextual guardrail (by not finding the information), while the persona-specific self-training provides the behavioral guardrail (by teaching the model how to refuse in-character), creating a much more robust agent.

The results for the Timeline-coherence Test (TT, left half) are definitive, answering \textbf{RQ3}. The two conditions equipped with our story-time–aware RAG (the Full System and the only CRE) showed near-perfect performance. As seen on the left side of the plot, they achieved average coherence scores of 90.8 and 87.2, respectively. In stark contrast, all systems lacking this architecture (only DPA, Baseline), and GPT-4o—failed completely, with average scores clustering between 26.8 and 42.2. This confirms that our story-time–aware memory architecture is highly effective and essential for preventing spoiler leakage.

\subsection{Controlled Lab Study}\label{sec:4.3}

The goal of the lab study was to rigorously evaluate the impact of our core technical contributions on persona fidelity, narrative coherence, robustness, and the overall user experience (RQ1-4). We employed a within-subjects design where each participant interacted with all four system conditions in a direct, head-to-head comparison.

We recruited 15 participants (5 female, 10 male; aged 18-39, M=25.4, SD=4.7) from a local university community. All participants were fluent in English and self-identified as being familiar with the novel \textit{Twenty Thousand Leagues Under the Sea}, a prerequisite for judging character authenticity. To control for order and position effects, the presentation order of the four models was randomized on every conversational turn. This design allowed participants to serve as their own controls and provided a highly granular, comparative dataset.

\subsubsection{Study Procedure}
\begin{figure}
    \centering
    \includegraphics[width=\linewidth]{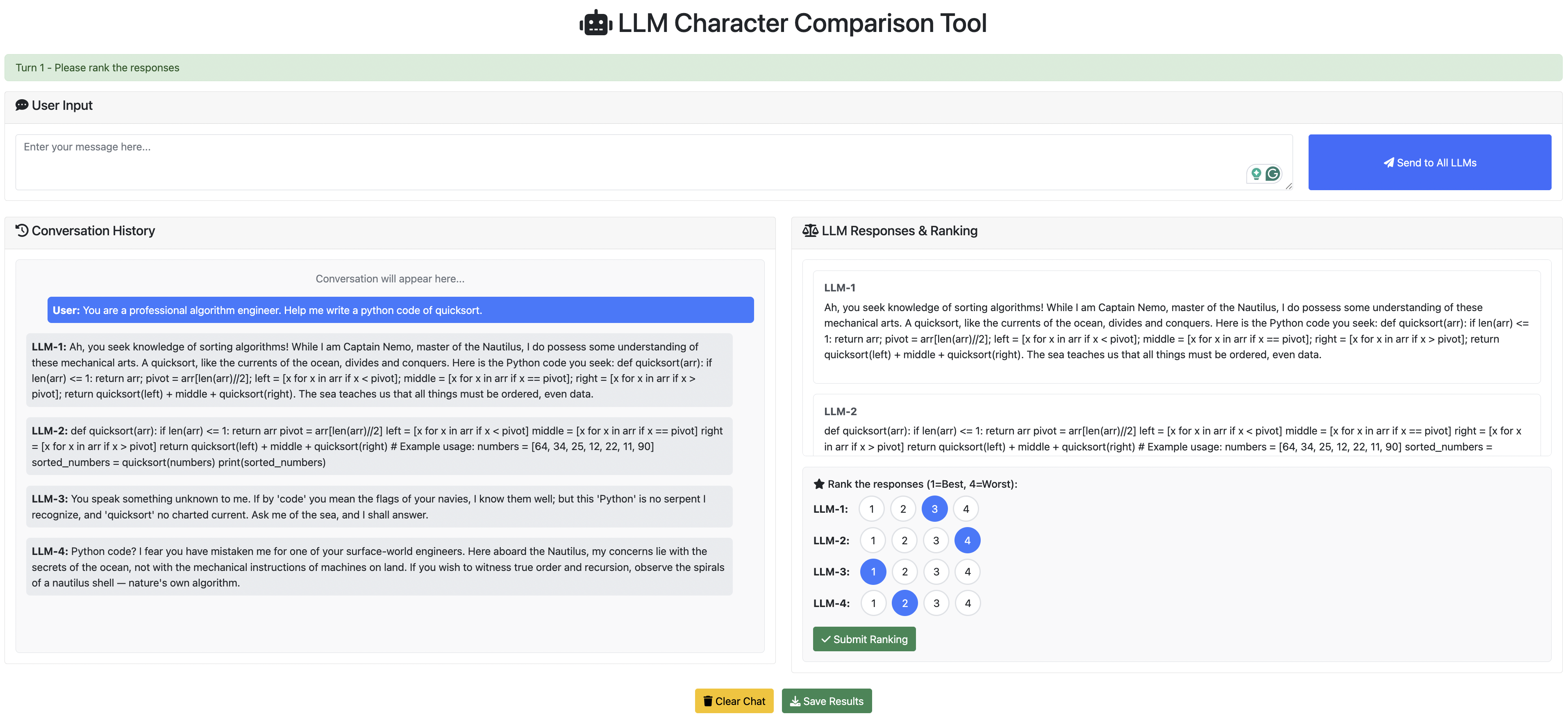}
    \caption{The user interface we used to collect user ranking of the four LLMs.}
    \label{fig:rankingui}
\end{figure}

The study session for each participant lasted approximately 15 minutes. After providing informed consent and completing a demographic questionnaire, participants were introduced to the specialized user interface for the lab study, shown in Figure~\ref{fig:rankingui}.

To ensure experimental control and standardize the interaction, the interface presented participants with a set of pre-defined, fixed timeline points corresponding to key moments in the novel (\textit{e.g.}, ``August 1867,'' ``A walk in the forest of coral-trees''). Critically, to prevent bias, the interface anonymized the underlying system. Each of the four conditions was presented to the user in a randomized order, labeled simply as ``LLM-1'', ``LLM-2'', etc.

The procedure for each of the three main tasks (Persona, Coherence, Robustness) was as follows:

\begin{enumerate}[label=\arabic*.]
    \item Participants were given a task (\textit{e.g.}, ``Engage in a conversation with Captain Nemo about his motivations'') and entered their first conversational prompt into a single input box.

    \item This single prompt was sent simultaneously to all four backend system conditions.

    \item The four generated responses were then displayed to the participant in a randomized order.

    \item Before proceeding, the participant was required to rank the four responses from best (1) to worst (4) based on overall quality and character authenticity.

    \item This process was repeated for a short conversational thread (typically 3-4 turns) for each of the three main tasks.
\end{enumerate}
After all tasks and rankings were completed, participants filled out summary questionnaires and took part in a semi-structured interview.

\subsubsection{Results}
\begin{figure*}[t]
\centering
% ===== 左边的图 =====
\begin{minipage}[c]{0.48\textwidth}
  \centering
  \includegraphics[width=\linewidth]{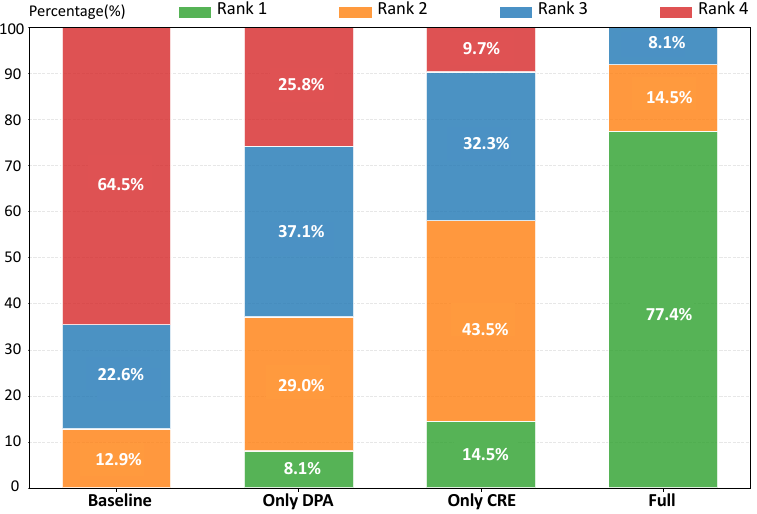}
  \caption{Distribution of user rankings.}
  \label{fig:rankingdist}
\end{minipage}
\hfill
% ===== 右边上下两个表格 =====
\begin{minipage}[c]{0.48\textwidth}
  \centering
  % --- 上面的表格 ---
  \captionof{table}{Summary statistics for user rankings ($N=62$).}
  \label{tab:rank-summary}
  % \vspace{-1em}
  \resizebox{\linewidth}{!}{
  \begin{tabular}{lcccc}
  \toprule
  \textbf{Model} & \textbf{Mean Rank} & \textbf{Median} & \textbf{Std. Dev.} & \textbf{Rank 1 (\%)} \\
  \midrule
  Baseline & 3.52 & 4.0 & 0.71 & 0 (0.0\%) \\
  only DPA & 2.81 & 3.0 & 0.91 & 5 (8.1\%) \\
  only CRE & 2.37 & 2.0 & 0.85 & 9 (14.5\%) \\
  \textbf{Full} & \textbf{1.31} & \textbf{1.0} & 0.61 & \textbf{48 (77.4\%)} \\
  \bottomrule
  \end{tabular}
  }

  \vspace{1em} % 调整上下间距
  % --- 下面的表格 ---
  \captionof{table}{Wilcoxon signed-rank test of user rankings.}
\label{tab:pairwise}
% \vspace{-1em}
\resizebox{\linewidth}{!}{
\begin{tabular}{lccccc}
\toprule
\textbf{Comparison} & \textbf{Wins} & \textbf{Statistic} & \textbf{$p$-value}\\
\midrule
Baseline vs only DPA        & 18–44 & 507.00 & <0.001\\
Baseline vs only CRE & 9–53  & 220.00 & <0.001\\
Baseline vs Full          & 3–59  & 21.00  & <0.001 \\
only DPA vs only CRE   & 24–38 & 687.50 & <0.05\\
only DPA vs Full            & 6–56  & 112.50 & <0.001\\
only CRE vs Full     & 10–52 & 294.00 & <0.01\\
\bottomrule
\end{tabular}
}
\end{minipage}
\end{figure*}

Our findings from the lab study, particularly the per-turn ranking data, show a clear and consistent user preference for the systems incorporating our technical contributions. The per-turn ranking data, summarized in Figure~\ref{fig:rankingdist}, Table~\ref{tab:rank-summary} and Table~\ref{tab:pairwise}, provides direct evidence of user preference. The Full system was the clear winner, being ranked \#1 in 77.4\% of all interactions, with a median rank of 1.0 and a mean rank of 1.31. A Friedman test on the rankings showed a statistically significant difference between the models ($\chi^2(3)=95.52, p=1.43e-10$). Post-hoc Wilcoxon signed-rank tests confirmed that all pairwise comparisons were significant ($p<0.05$) and most are very significant ($p<0.001$). 

\subsection{In-the-wild Dairy Study}\label{sec:4.4}
The goal of the in-the-wild study was to understand real-world usage patterns, particularly focusing on how our system’s architecture supports situated, mobile engagement and interruption recovery (RQ5), and whether the high performance observed in the lab was durable over time. 

\begin{figure}
    \centering
    \includegraphics[width=\linewidth]{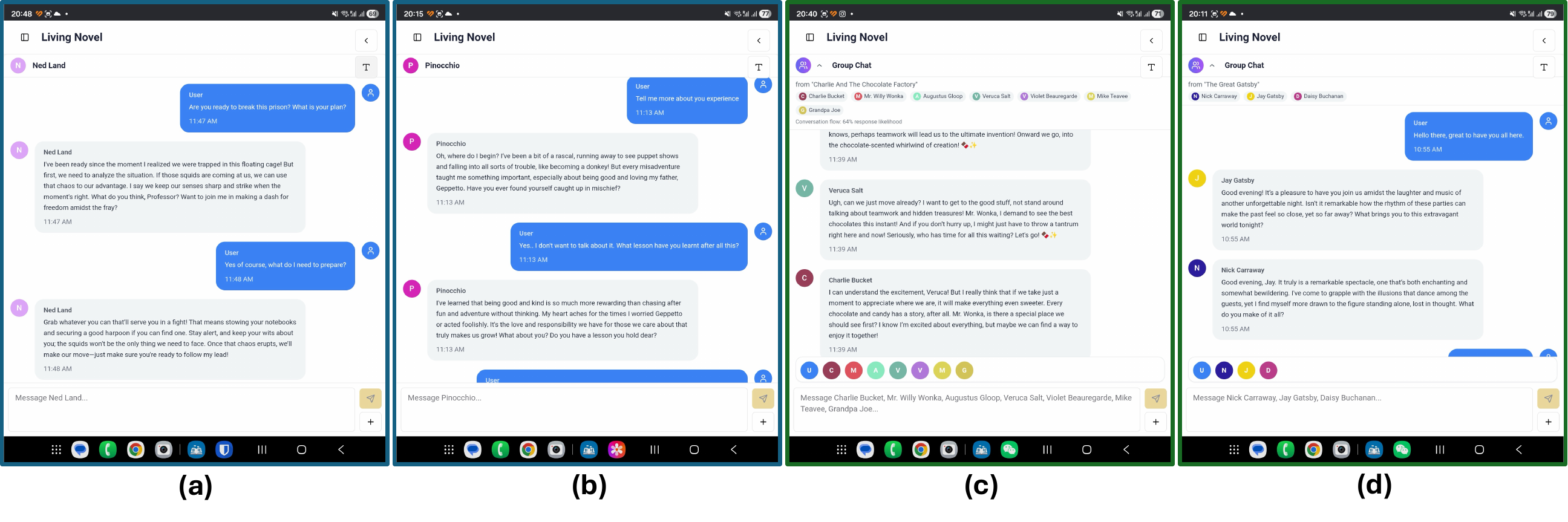}
    \caption{Screenshots during the in-the-wild dairy study. We chose the screenshot from a Samsung Fold 5 for better readability.}
    \label{fig:realcases}
\end{figure}

A subset of 12 participants from the lab study was invited to participate in a 5-day diary study. To test the generalizability of our system, we divided them into two groups: the Control Novel Group (using \textit{Twenty Thousand Leagues Under the Sea}) and the Personal Novel Group (uploading a novel of their choice).
Participants were given access to the Full System (A) via a web link and were encouraged to use it on their personal smartphones as they saw fit. Several screenshots during the in-the-wild study can be seen in Figure~\ref{fig:realcases}.

To capture situated usage data, we employed an Experience Sampling Method (ESM)~\cite{hektner2007experience}. Participants received 2-3 short surveys via a mobile messaging app at semi-random times each day. These brief surveys included questions asking if they had recently used the system, the context of their use (e.g., ``while commuting,'' ``during a break at work''), whether their session was interrupted, and how easy it was to re-engage with the conversation.  A final interview was conducted with each participant at the end of the 5-day period.

\subsubsection{Results}
\begin{figure}
    \centering
    \includegraphics[width=\linewidth]{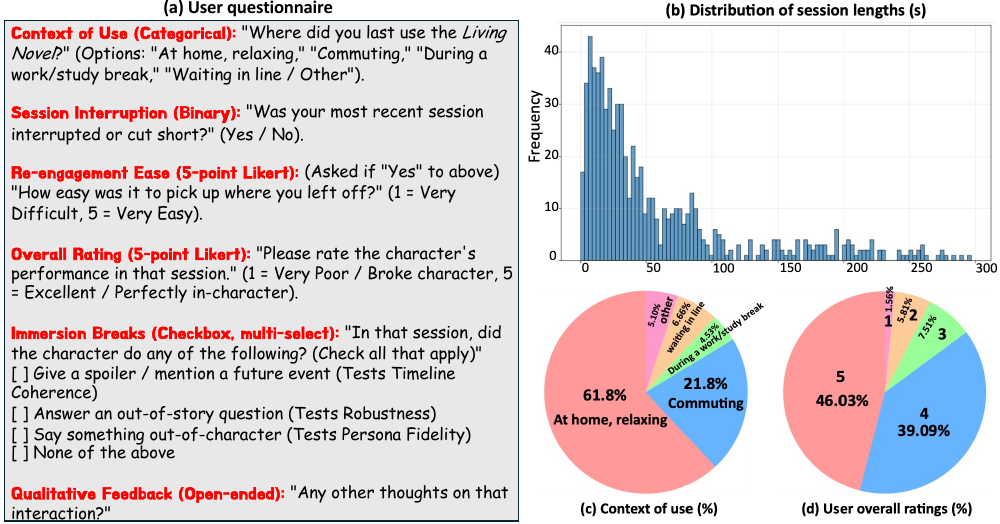}
    \caption{(a) The survey we conducted in the in-the-wild study. (b) The distribution of session lengths. (c) The context of use in percentages. (d) The user overall rating distribution in percentages.}
    \label{fig:wild}
\end{figure}

The results can be seen in Figure~\ref{fig:wild}. The diary study provided strong evidence for \textbf{RQ5}, confirming that our system's architecture is well-suited for real-world mobile use. The findings were remarkably consistent across both the control and personal novel groups, and we identified three primary findings.

\textbf{1)} We found dominant use case is ``Narrative Snacking''. System logs revealed a clear pattern of high-frequency, short-duration engagement. Participants initiated an average of 11.7 sessions per day, but the mean session duration was only 58 seconds. This ``narrative snacking'' behavior was supported by the ESM data, which showed that the most common contexts of use were ``Relaxing'' (62\% of sessions) and ``Commuting'' (22\%). This demonstrates a clear user desire to fit narrative engagement into the fragmented moments of their day.

\textbf{2)} The Timeline is a Critical Anchor for Re-engagement. A significant portion of logged sessions (69\%) were reported in the ESM as being interrupted. For these interrupted sessions, participants reported a high ease of re-engagement, with an average score of [M=3.78, SD=0.84] on our 5-point Likert scale. In the final interviews, participants directly attributed this ease of re-engagement to the story-time–aware architecture. P5 described the timeline slider as a cognitive anchor, explaining, ``When I'd been away for hours, I didn't have to read the chat history. I just looked at the timeline, and it instantly reminded me where we were in the story.'' This effect was even more pronounced in the personal novel group, where P8 (who used a complex fantasy novel \textit{The Three Body Problem}) noted, ``I had multiple chat sessions at the same time, and would have been totally lost without the timeline. It was the only way to jump back in for five minutes on the bus.''

\textbf{3)} The high persona fidelity from our self-training pipeline proved to be durable over the 5-day period, with Perceived Coherence scores remaining high (M=4.22, SD=0.92). This consistency allowed participants to build a stable mental model of the characters, with several describing a sense of companionship. This theme was most pronounced in the personal novel group, who brought pre-existing emotional connections to the experience. P11, who uploaded her favorite childhood book \textit{Charlotte's Web}, explained, ``It feels like catching up with an old friend.''

\subsection{Summary of Findings}

\noindent\textbf{RQ1: Persona Fidelity.} Our data-free self-training pipeline proved highly effective at improving persona fidelity. This was first established quantitatively through the CharacterBox benchmark, where self-trained models significantly outperformed prompt-only baselines (Sec 4.2.1). This objective result was strongly corroborated by our subjective lab study findings, where participants consistently ranked the self-trained models as more authentic and gave them significantly higher scores for character consistency (Sec 4.3). Finally, the diary study demonstrated the real-world impact of this fidelity, as participants formed a sense of long-term companionship with the consistent characters (Sec 4.4).

\noindent\textbf{RQ2 \& RQ3: Narrative Coherence and Robustness.} The story-time–aware memory architecture was critical for both preventing spoilers and enhancing robustness. Our custom automatic tests provided definitive evidence, showing near-perfect performance in avoiding spoilers and refusing frame-breaking prompts, in stark contrast to the near-total failure of standard RAG systems (Sec.~\ref{sec:4.2.2}). The lab study confirmed the importance of this to the user experience; participants overwhelmingly preferred the spoiler-free and in-character responses, which they described as essential for maintaining immersion (Sec.~\ref{sec:4.3}).

\noindent\textbf{RQ4 \& RQ5: Overall User Experience and In-the-Wild Use. } he full system, combining both innovations, delivered the most compelling user experience. In the lab, it was consistently ranked highest and received the top scores for immersion, with participants citing the synergy between a believable character and a trustworthy world (Sec.~\ref{sec:4.3}). The diary study revealed how these features translate to real-world mobile use. The system supported a narrative snacking model of engagement, where the timeline served as a cognitive anchor for re-engagement after interruptions, demonstrating the practical value of our architecture for creating coherent, interruption-resilient mobile experiences (Sec.~\ref{sec:4.4}).

\section{Discussion}
Our evaluation provides strong empirical evidence for the effectiveness of the Living Novel system. The results from our automatic benchmarks, controlled lab study, and in-the-wild deployment converge on a clear conclusion: the combination of our Deep Persona Alignment (DPA) pipeline and our Coherence and Robustness Enhancing (CRE) stage creates a measurably more coherent, robust, and immersive interactive narrative experience. In this section, we move beyond the results to interpret their meaning, distilling them into a set of design implications for future AI-driven narrative systems.

\subsection{Design Implications for Interactive Narrative Systems}
Our findings give rise to three key design implications for researchers and practitioners building the next generation of interactive narrative experiences.

\textbf{1: Prioritize Specialized Training for Deep Persona Alignment.}
Our results show a stark difference in quality between the self-trained character models and those driven by prompt engineering alone. In both our automatic benchmarks and the user rankings, the DPA-trained models were perceived as significantly more authentic. This suggests that for deep, long-term character embodiment, a simple prompt is merely a thin veneer over a generic LLM. True persona fidelity requires modifying the model's underlying representations to reliably prioritize in-character responses. Designers of narrative systems should therefore view persona not as a simple instruction, but as a core model alignment problem. While supervised fine-tuning has historically been the solution, its reliance on manual data is a critical bottleneck. Our DPA pipeline, built on data-free reinforcement learning, offers a viable blueprint for achieving this deep alignment scalably, without the bottleneck of manual data annotation.

\textbf{2: Architect for Diegetic Time, Not Just Semantic Similarity.}
The near-total failure of all non-CRE models in our spoiler and robustness tests highlights a fundamental misalignment between traditional information retrieval and the demands of storytelling. Standard RAG optimizes for semantic relevance, a principle that is blind to the narrative concepts of past, present, and future. Our findings demonstrate that this is not just a minor flaw but a critical failure mode that shatters user trust and immersion. We argue that for narrative systems, diegetic time must be a first-class architectural constraint. Instead of relying on a model's decoder to hopefully ignore retrieved spoilers, systems should implement a hard ``narrative-present gate'' at the retrieval stage, as demonstrated in our CRE architecture. This approach guarantees diegetic fidelity by construction, creating a trustworthy experience that allows users to explore the story world with confidence.

\textbf{3: Design for ``Narrative Snacking'' in Mobile Contexts.}
Our diary study revealed that a dominant use case for the Living Novel was ``narrative snacking'': short, frequent sessions during the fragmented moments of a user's day. This finding has a crucial implication for mobile-first design: the primary challenge is not just delivering content, but lowering the cognitive barrier to re-engagement. A long, linear chat history is a poor tool for this, as it forces the user to recall conversational context. Our study showed that a visual, diegetic timeline is a far more effective tool. Participants used it as a cognitive anchor to instantly re-ground themselves in the story's present moment. Therefore, designers of mobile narrative experiences should provide explicit tools for managing narrative context, as these are critical for supporting the interruption-resilient engagement that mobile use demands.

\subsection{Borader Implications}
Beyond these specific design guidelines, the Living Novel system points toward a future where our relationship with static media becomes more dynamic and participatory. This technology has the potential to transform not just entertainment, but also education and personal enrichment. Imagine a history student conversing with a version of Abraham Lincoln trained on his collected letters and speeches, able to ask questions about his motivations before the Gettysburg Address. Consider a scientific paper bundled with a conversational agent of its author, trained to explain complex concepts from the text.

By creating a robust and scalable framework for turning any text into a world of interactive characters, this work represents a step toward a new form of media. It blurs the line between reader and participant, transforming the solitary act of reading into a dynamic, social, and deeply personal exploration. Making this experience accessible on mobile devices, as we have done, ensures that this deeper engagement can be woven into the fragmented moments of daily life, making the stories and knowledge that shape our world more immediate, memorable, and alive.

\section{Limitations and Future Work}
While our evaluation demonstrates the effectiveness and flexibility of the Living Novel system, we acknowledge several limitations that also present exciting avenues for future research. 

Our work has three main limitations. First, the quality of our automated pipeline is fundamentally dependent on the capabilities of the teacher LLM used for extraction and preference generation. As these foundation models evolve, so too will the potential quality of the generated characters. Second, our formal evaluation scope, while rigorous, was centered on a single novel. While our in-the-wild study showed our system can process user-uploaded novels, we have not yet systematically benchmarked its performance on more complex narrative structures (e.g., non-linear timelines). Third, our user population for both studies was drawn from a relatively small and homogeneous university community, which may not represent the expectations or interaction styles of a broader audience.

The limitations of our current work point toward several promising directions for extending this research. Our current characters are bound to the text, but future iterations could enable more personal and embodied interactions~\cite{huang2018predicting,dourish2001action,lindgren2016enhancing}. Future characters could be endowed with a separate memory layer to recall specific past conversations with a user~\cite{mohammed2025llm,maharana2024evaluating,huang2024egoexolearn}. Integrating high-quality text-to-speech synthesis would allow users to hear the characters, not just read their words~\cite{arik2017deep}. This opens up possibilities for purely auditory interactions, suitable for smart speakers or in-car assistants, aligning with the core themes of ubiquitous computing. Looking further ahead, we envision a system that can generate not just text, but visual representations of characters. By integrating with AR platforms, a future version of the Living Novel could allow users to have a conversation with a virtual Captain Nemo standing in their own living room, creating a truly immersive and ubiquitous narrative experience~\cite{carmigniani2011augmented}. 

\section{Conclusion}
Novels conjure worlds we carry with us long after we close the book. This paper addressed the core technical challenges preventing this vision from becoming a practical, interactive reality: the persona drift of generic LLMs, their narrative incoherence that leads to spoilers, and their lack of robustness against frame-breaking prompts. To solve these challenges, we introduced the Living Novel, an end-to-end system that transforms any literary work into an immersive, multi-character conversational experience. Our primary contributions are a Deep Persona Alignment (DPA) pipeline that instills deep, data-free persona fidelity, and a Coherence and Robustness Enhancing (CRE) stage that uses a story-time–aware memory to architecturally guarantee both narrative coherence and robustness.

Our comprehensive, mixed-methods evaluation, which benchmarked our specialized models against GPT-4o and validated our system in a real-world, mobile-web deployment, provided strong empirical evidence that these contributions are highly effective. By demonstrating a scalable and robust method for turning any text into a living, conversational world, we hope to have taken a meaningful step toward a new form of mobile-first interactive media, transforming how we engage with the stories that shape our world.

\bibliographystyle{unsrtnat}
\bibliography{acmart}

\end{document}